\documentclass[conference]{IEEEtran}
\usepackage{graphicx}%------------------------------added by ZY
\usepackage{subfigure}
\usepackage{float}%---------------------------------added by ZY
\usepackage{algorithm}%-----------------------------added by ZY
\usepackage{algorithmic}%---------------------------added by ZY
\usepackage{mathrsfs}%------------------------------added by ZY
\usepackage{amsfonts}
\usepackage{array}
\usepackage{amssymb,amsmath}
\usepackage{longtable}
\usepackage{rotating}
\usepackage{multirow}
\usepackage{epstopdf}
\usepackage{cite}
\usepackage{float}
\usepackage[marginal]{footmisc}
\usepackage{stfloats}
\usepackage[colorlinks,
            linkcolor=black,
            anchorcolor=black,
            urlcolor=blue,
            citecolor=black]{hyperref}
\usepackage[hyphenbreaks]{breakurl}

\renewcommand{\baselinestretch}{1}

\def\BibTeX{{\rm B\kern-.05em{\sc i\kern-.025em b}\kern-.08em
    T\kern-.1667em\lower.7ex\hbox{E}\kern-.125emX}}

\renewcommand{\baselinestretch}{1}
\begin{document}

\title{Resource Allocation for Intelligent Reflecting Surface Aided Cooperative Communications}
\author{\IEEEauthorblockN{Yulan Gao\textsuperscript{1,2}, Chao Yong\textsuperscript{1}, Zehui Xiong\textsuperscript{2}, Dusit Niyato\textsuperscript{2},  Yue Xiao\textsuperscript{1}, Jun Zhao\textsuperscript{2}}

\IEEEauthorblockA{\textsuperscript{1}{The National Key Laboratory of Science
and Technology on Communications} \\
{University of Electronic
Science and Technology of China, Chengdu 611731, China }\\
{email: xiaoyue@uestc.edu.cn}}

\IEEEauthorblockA{\textsuperscript{2}{The School of Computer Science and Engineering,}
{Nanyang Technological University, Singapore  639798}}
}

\maketitle

\begin{abstract}
This paper investigates an intelligent reflecting surface (IRS) aided cooperative communication network, where the IRS  exploits large reflecting elements to proactively steer the incident radio-frequency wave towards destination terminals (DTs).
As the number of reflecting elements increases, the reflection resource allocation (RRA) will become urgently needed in this context, which is due to the non-ignorable energy consumption.
The goal of this paper, therefore, is to realize the RRA besides the active-passive beamforming design, where RRA is based on the introduced modular IRS architecture.
The modular IRS consists with multiple modules, each of which has multiple reflecting elements and is equipped with a smart controller,  all the controllers can communicate with each other in a point-to-point fashion via fiber links.  Consequently, an optimization problem is formulated to maximize the minimum SINR at DTs, subject to the module size constraint and both individual source terminal (ST) transmit power and the reflecting coefficients constraints.
Whereas this problem is NP-hard due to the module size constraint, we develop an approximate solution by introducing the mixed row block $\ell_{1,F}\text{-norm}$ to transform it into a suitable semidefinite relaxation.
Finally, numerical results demonstrate the meaningfulness of the introduced modular IRS architecture.
\end{abstract}
\IEEEpeerreviewmaketitle

\section{Introduction}
%\IEEEPARstart{C}{ooperative} communication is a key technique adopted in current wireless systems and applications, such as long-term evolution (LTE), LTE-Advanced cellular systems, and WLAN, to improve quality of service, coverage, and resource usage efficiently.

\IEEEPARstart{O}{wing} to the highly demanding of forthcoming and future wireless networks fifth generation (5G and beyond), the biggest challenge in the wireless industry today is to meet the soaring demand at the cost of resulting power consumption \cite{Grassi2017Uplink, NGMN2015}.
For instance, for mMIMO, adopting a higher amount of base station antennas to serve multiple users concurrently not only entails the increased radio frequency chains and maintenance cost, but also significantly decreases the overall performance level.
Therefore, addressing this issue means introducing innovation technologies in future/\mbox{beyond-5G} wireless networks, which are spectral-energy efficient and cost-effective \cite{Zenzo2019Wireless, Wu2019Towards}.
Recently, as a remedy to balance the spectral-energy and cost efficiency, communication systems employing intelligent reflecting surface have been emerged as a promising paradigm to provide communication services via exploiting large software-controlled reflection elements \cite{Gao2020Reconfigurable}.
The intelligent reflecting surface (IRS) provides a new degree of freedom to further enhance the wireless link performance via proactively steering the incident radio-frequency wave towards destination terminals (DTs) as its important feature.

The IRS-aided communications refer to the scenario that a large number of software-controlled reflecting elements  with adjustable phase shifts for reflecting the incident signal.
%As such,  the phase shifts of all reflection elements can be tuned adaptively according to the state of networks, e.g., the channel conditions and the incident angle of the signal by the source terminal (ST).
%Notably, different from the conventional half and full-duplex modes, in IRS-aided communications, the propagation environment can be improved without incurring additional noise at the reflector elements.
Currently, considerable research attention has been paid for IRS-aided communications \cite{ Wu2018Intelligent,  Huang2019Reconfigurable}.
%Among the early contributions in this area, \cite{Wu2019Towards, Zenzo2019Reconfigurable} summarized the main communication applications and competitive advantages of the IRS technology.
The common assumption in the existing studies for IRS-aided communications is that all the reflecting elements are used to reflect the incident signal, i.e., adjusting reflecting coefficient of each meta-element simultaneously each time.
However, along with the use of a large number of high-resolution reflecting elements, especially with continuous phase shifters, activating all the reflecting elements every time may result in significant  power consumption and increasing implementation complexity.
%Moreover, the hardware support for the  IRS implementation is the use of a large number of tunable metasurfaces.
%Specifically, the tunability feature can be realized by introducing mixed-signal integrated circuits (ICs) or diodes/varactors, which can  vary both the resistance and reactance, offering complete
%local control over the complex surface impedance \cite{Liu2019Intelligent,Tan2018Enabling}.
%According to the IRS power consumption model presented in \cite{Huang2019Reconfigurable} and the hardware support, triggering the entire IRS not only incurs increased power consumption, but also entails the increased latency of adjusting phase-shift and accelerates equipment depreciation.
Therefore, implementing RRA is significantly important for IRS-aided communications.

In this paper, we consider the cooperative communication network in which multiple single-antenna source terminals (STs) reach the corresponding single-antenna DTs through an IRS that forwards a suitably phase-shifted version of the transmitted signal.
Our goal is to maximize the minimum signal-to-interference-plus-noise ratio (SINR) at DTs  via joint RRA, transmit power allocation, and the corresponding passive beamforming design.
%Specifically, the novelty and contributions of this paper mainly lie in the following aspects.
We develop a partially controllable modular IRS structure that  divides all the reflecting elements into multiple modules, each of which is attached with a smart controller, and all the modules can be independently controlled and communicated with each other via fiber links.
Furthermore, from an operational standpoint, independent module activation can be implemented easily.
Inspired by \cite{Mehanna2013Joint}, the RRA can be realized via module activation, which is based on the proposed modular architecture of IRS.
Specifically, we formulate the max-min SINR problem to optimize the modules activation, transmit power allocation, and the corresponding passive beamformer design under maximum transmit power per ST and module size constraints. To the best of our knowledge, this is the first work that studies the RRA via the modules activation.
To deal with the joint optimization problem, we transform the hard module size constraint into the group sparse constraint by introducing the mixed row block $\ell_{1,F}\text{-norm}$ \cite{Mehanna2013Joint}, which yields a suitable semidefinite relaxation.
Consequently, the convex approximate problem of the original max-min problem can be developed, which can be solved by the existing convex optimization solver such as CVX.

The reminder of this paper is organized as follows. The partially controllable modular IRS architecture,  modules activation, and the hard module size constraint convex relaxation are presented in Section II.
Section III describes the modules activation and minimum SINR maximization.
Section IV reports numerical results that are used to assess the effectiveness of the approximate solution and the meaningfulness of the modular IRS.
Conclusions are presented in Section V.

Matrices and vectors are denoted by bold letters. ${\mathbf I}_N$, ${\mathbf 0}_N,$ and ${\mathbf e}_n$ are the $N\times N$ identity matrix, the $N\times 1$ all-zero column vector, and the $N\times 1$  elementary vector with a one at the $n \text{th}$ position, respectively.
${\mathbf A}^{T}, {\mathbf A}^{\dag},$ ${\mathbf A}^{-1}$, and $||{\mathbf A}||_F$ denote transpose, Hermitian, inverse, and Frobenius norm of matrix ${\mathbf A},$ respectively.
$\text{Re}(\cdot)$, $\text{Im}(\cdot),$ and  $|\cdot|$  denote real part, imaginary part, and modulus of the enclosed vector, respectively.

\section{System Model and Problem Formulation }

\subsection{System Model}
We consider a two-hop network where there are $K$ S-D pairs, each pair communicating through the IRS.
Each user terminal is equipped with a single antenna.
%As shown in Fig. \ref{fig:1}, we consider a two-hop link of slowly-varying P2P network where an IRS is adjoined to $K$ user pairs,  where a user pair includes one ST and one DT, and each user terminal is equipped with a single antenna.
%Let ${\cal S}:=\{s_1, s_2, \ldots, s_K\}$ and ${\cal D}:=\{d_1, d_2, \ldots, d_K\}$ be the sets of STs and DTs, respectively.
The index set of user pairs is denoted by ${\cal K}:=\{1, 2, \ldots, K\}.$
%The modular architecture of the IRS is shown in Fig. \ref{fig:2} for which multiple parallel controlled switches (on/off) are considered. The setting can can be regarded as generalization of the IRS architecture introduced in \cite{Wu2019Towards}.
For the partially controlled IRS, the total $N$ reflecting elements are divided into $M$ modules, each of which is attached with a smart controller, each module consists of $L$ elements, and $N=ML.$
Define ${\cal M}:=\{1, 2, \ldots, M\}$ as the index set of reflection modules.
%Consequently, RRA of IRS can be realized by the modular of reflecting elements and the parallel controllers design.
The channels of two-hop communications are assumed to experience quasi-static block fading, i.e., the channel coefficient from the STs to the IRS and the IRS to the DTs remain constant during each time slot, but may vary from one to another \cite{Feng2013Device}.
Let ${\mathbf h}_{k,m}\in{\mathbb C}^{L\times 1 }$ and ${\mathbf g}_{m,k}\in{\mathbb C}^{L\times 1}$ denote the uplink channel vector from ST $k$ to the $m\text{th}$ module of IRS
and the downlink channel vector from reflection module $m$ to DT $k$, respectively.
The associated passive beamformer at the $m\text{-th}$ module of IRS denoted by ${\pmb\Phi}^m=\text{diag}[\phi_{(m-1)L+1}, \ldots, \phi_{(m-1)L+l}, \ldots, \phi_{mL}]\in{\mathbb C}^{L\times L},$ where $\phi_{(m-1)L+l}$ is the $l\text{th}$ entry.
We assume that all the modules can potentially serve the STs' transmitting.
Note that if all modules are activated to serve the ST-DT communications, the problem becomes a special case which is simpler to solve.
The passive beamformer at IRS denoted by $\pmb\Phi\in{\mathbb C}^{ N\times N},$ and the associated channel from ST $k$ to the IRS and the downlink channel from the IRS to DT $k$, denoted by ${\mathbf h}_k\in{\mathbb C}^{N\times 1}$ and ${\mathbf g}_k\in{\mathbb C}^{N\times 1},$ respectively, are expressed as
\begin{subequations}\label{s:1}
\begin{align}
{\pmb\Phi}&=\text{diag}\{{\pmb\Phi}^1, {\pmb\Phi}^2, \ldots, {\pmb\Phi}^M\},\\
{\mathbf h}_k&=[ ({\mathbf h}_{k,1})^T, ({\mathbf h}_{k,2})^T, \ldots, ({\mathbf h}_{k,M})^T ]^T, \forall k\in{\cal K},\\
{\mathbf g}_k&=[({\mathbf g}_{1, k})^T, ({\mathbf g}_{2,k})^T, \ldots, ({\mathbf g}_{M,k})^T]^T, \forall k\in{\cal K}.
\end{align}
\end{subequations}
%In the following analysis--for the sake of simplicity--we consider that the reflection coefficient is peak-power constrained \cite{Guo2019Weighted}: ${\cal X}\triangleq\{\phi_{(m-1)L+l}: |\phi_{(m-1)L+l}|\leq 1, \forall m=1, 2,\ldots, M, l=1, 2, \ldots, L\}.$

The signal received at DT $k$ via IRS-aided link is expressed by
\begin{equation}\label{s:2}
\begin{aligned}
y_{k}&={\mathbf g}_{k}^{\dag}{\pmb\Phi}\sum_{k=1}^K\sqrt{p_{k}}{\mathbf h}_{k}z_k+u_k\\
&={\mathbf g}_{k}^{\dag}{\pmb\Phi}\sqrt{p_k}{\mathbf h}_{k}z_k+{\mathbf g}_{k}^{\dag}{\pmb\Phi}\sum_{j=1, j\neq k}^K\sqrt{p_{j}}{\mathbf h}_{j}z_j+u_k,
\end{aligned}
\end{equation}
where $z_k$ and $p_k$ represent the data symbol of ST $k$ and power, respectively,
$u_k\sim {\cal C}{\cal N}(0, \sigma^2)$ is the thermal noise experienced by DT $k$ and the second term accounts for the interference experienced by user pair $k$ from other user pairs $j\in{\cal K}, j\neq k.$
Then, the SINR achieved at DT $k$ is expressed as
\begin{equation}\label{eq:2}
\text{SINR}_k=\frac{p_k|{\mathbf g}_k^{\dag}{\pmb\Phi}{\mathbf h}_k|^2}
{\sum_{j=1,j\neq k}^Kp_j|{\mathbf g}_k^{\dag}{\pmb\Phi}{\mathbf h}_j|^2+\sigma^2}, \forall k\in{\cal K}.
\end{equation}

\subsection{ Modules  Activation}
Here, we develop the optimization problem to maximize the minimum SINR at DTs, subject to the module size constraint and both maximum transmit power per ST and reflecting coefficient constraints.
Different from the previous studies that mainly focused on designing the appropriate transmit beamforming at the STs and passive beamforming at IRS, in this paper, we design the modules activation at the IRS besides the active-passive beamforming optimization.
As aforementioned, in the modular IRS architecture, all reflecting elements are divided into modules which are activated by multiple controllers in parallel.
%In the proposed modular IRS architecture, the reflecting elements management can be implemented instead of tuning all the entries at IRS each time.
Now suppose that only $Q\leq M$ modules are available, and thus only $QL$ reflecting elements can serve the STs simultaneously.
Inspired by \cite{Mehanna2013Joint}, the design problem is to jointly select the best $Q$ out of $M$ modules, and designing the transmit power $\{p_k\}_{k=1}^K$ and the corresponding beamformer at the IRS so that the minimum SINR among DTs is maximized, subject to the maximum transmit power constraint and reflecting coefficient.

Define the $N\times 1$ vector ${\pmb\phi}:=[({\pmb\phi}^1)^T, ({\pmb\phi}^2)^T, \ldots, ({\pmb\phi}^M)^T]^T$, where ${\pmb\phi}^m:=[\phi_{(m-1)L+1}^{\dag}, \ldots, {\phi}_{mL}^{\dag}]^T\in{\mathbb C}^{L\times 1}$ is the $m\text{th}$ block of vector ${\pmb\phi}$, $ \forall m=1, 2, \ldots, M.$
Denote the $M\times1 $ vector $\bar{\pmb\phi}:=[||{\pmb\phi}^1||_2, ||{\pmb\phi}^2||_2, \ldots, ||{\pmb\phi}^M||_2]^T.$
If module $m$ is activated, vector ${\pmb\phi}^m$ must be set to zero, consequently, $||{\pmb\phi}^m||_2=0.$
Hence, the maximization of minimum SINR among DTs problem via joint module activation and active-passive beamformer design can be expressed by
\begin{align}%\label{eq:3}
\text{(P0)}& \max_{\left\{{\pmb\Phi}, \{ p_k\}_{k=1}^K\right\}} \min_{k}~ \text{SINR}_k \label{eq:3-1}\\
\text{s. t.~} &||\bar{\pmb\phi}||_0\leq Q,  \label{eq:3-2}\\
&p_k\leq p_k^{\max}, \forall k\in {\cal K}\label{eq:3-3}\\
&|\phi_n|\leq 1, \forall n=1, 2, \ldots, N,\label{eq:3-4}
\end{align}
where the $\ell_{0}\text{-norm}$ denotes the number of activated modules, i.e., $||\bar{\pmb\phi}||_0:=\left|\{m: ||{\pmb\phi}^m||_2\neq 0 \}\right|$,  and  $Q\leq M$ is the upper bound of this number.
Note that (P0) is an NP-hard problem due to the module size constraint, and solving (P0) requires an exhaustive combinatorial search over all $\binom{Q}{M} $ possible patterns of $\tilde{\pmb\phi}.$
Thus, in the following, we aim to develop computationally efficient method to obtain a sub-optimal solution.

\subsection{Convex Relaxation and Problem Formulation}
%By the ${\ell}_{0,2}\text{-quasi}$ norm presented in \cite{Mehanna2013Joint},  (\ref{s:6}b) is equivalent to $||{\pmb\phi}||_0=\left|\left\{m: ||{\pmb\phi}_m||\neq 0  \right\}\right|\leq Q,$
%which means if the $m\text{th}$  resource block of IRS is not active, the $m\text{th}$ block in ${\pmb\phi}$ is zero, i.e., ${\pmb\phi}_m={\mathbf 0}.$
%Unfortunately, due to the IRS block constraint (\ref{s:6}b), solving (P1--1) requires an exhaustive combinatorial search over all $\binom{Q}{M} $ possible sparsity patterns of ${\pmb\phi},$  where the NP-hard problem (\ref{s:6}) must be solved for each of these patterns.

Define ${\mathbf A}_{k}=\text{diag}[{\mathbf g}_k^{\dag}]\in {\mathbb C}^{N\times N},$
the $N\times 1$ vector $\bar{\mathbf h}_{j, k}={\mathbf A}_{k}{\mathbf h}_{j},$
and the $(KN)\times 1$ vector $\bar{\mathbf H}^k=[(\bar{\mathbf h}_{1,k})^T, (\bar{\mathbf h}_{2, k})^T, \ldots,(\bar{\mathbf h}_{k,k})^T, \ldots, (\bar{\mathbf h}_{K,k})^T]^T.$
Thus, the expression of $\text{SINR}_k$ in (\ref{eq:2}) can be rewritten as
\begin{equation*}\label{s:13}
\begin{aligned}
\text{SINR}_{k}&=\frac{p_{k}{\pmb\phi}^{\dag}\bar{\mathbf h}_{k,k}\bar{\mathbf h}_{k,k}^{\dag}{\pmb\phi}}
{\sigma^2+\sum_{j=1, j\neq k}^Kp_{j}{\pmb\phi}^{\dag}\bar{\mathbf h}_{j,k}\bar{\mathbf h}_{j,k}^{\dag}{\pmb\phi}}
\end{aligned}
\end{equation*}
\begin{equation}\label{s:13}
\begin{aligned}
&=\frac{\bar{\pmb\phi}_k^{\dag}\bar{\mathbf h}_{k,k}\bar{\mathbf h}_{k,k}^{\dag}\bar{\pmb\phi}_k}
{\sigma^2+\sum_{j=1, j\neq k}^K\bar{\pmb\phi}_j^{\dag}\bar{\mathbf h}_{j,k}\bar{\mathbf h}_{j,k}^{\dag}\bar{\pmb\phi}_j},
\end{aligned}
\end{equation}
where $\bar{\pmb\phi}_k=\sqrt{p_k}{\pmb\phi}, \forall k\in{\cal K}. $
Define the $N\times K$ matrix $\bar{\pmb\Phi}=[\bar{\pmb\phi}_1, \bar{\pmb\phi}_2, \ldots, \bar{\pmb\phi}_K].$
In order to deal with the module size constraint (\ref{eq:3-2}) in (P0), similar to the \textit{group-sparsity inducing norm} $\ell_{1,2}\text{-norm}$ of vectors, we define the mixed convex norm $\ell_{1,F}$ of matrix as
%By introducing the mixed ${\ell}_{1, 2}\text{-norm}$, which was first presented in the context of the group \textit{least-absolute selection and shrinkage operator} (group \textit{Lasso}) \cite{Yuan2006Model}, the triggered module size can be effectively appropriated by replacing the ${\ell}_{0,2}\text{-norm}$ with ${\ell}_{1, 2}\text{-norm}$, i.e.,
%$||{\pmb\phi}||_{1,2}\triangleq \sum_{m=1}^M ||{\pmb\phi}^m ||_2.$
%Moreover, similar to the \textit{group-sparsity inducing norm} $\ell_{1,2}\text{-norm}$ of vectors, we define the mixed convex norm $\ell_{1,2}$ of matrix \cite{Lin2016Joint} as
\begin{equation}\label{eq:4}
||\bar{\pmb\Phi}||_{1,F}=\sum_{m=1}^M||\bar{\pmb\Phi}^m||_F,
\end{equation}
where $\bar{\pmb\Phi}^m\in{\mathbb C}^{L\times K}$ represents the $m\text{th}$ row block of matrix $\bar{\pmb\Phi},$ i.e., $\bar{\pmb\Phi}^m=[\sqrt{p_1}{\pmb\phi}^m, \ldots, \sqrt{p_K}{\pmb\phi}^m], \forall m\in{\cal M}.$
Notably, the mixed $\ell_{1,F}\text{-norm}$ of matrix $\bar{\pmb\Phi}$, which implies that each $||\bar{\pmb\Phi}^m||_2$ (or equivalently ${\pmb\phi}^m$) is encouraged to be set to zero, therefore inducing group-sparsity.
Thus, instead of using the hard module size constraint (\ref{eq:3-2}), the $\ell_{1,F}\text{-norm}$ can be employed to promote sparsity, leading to
\begin{align}
\text{(P1)}~&\max_{\bar{\pmb\Phi}}\min_k ~\frac{\bar{\pmb\phi}_k^{\dag}\bar{\mathbf h}_{k,k}\bar{\mathbf h}_{k,k}^{\dag}\bar{\pmb\phi}_k}
{\sigma^2+\sum_{j=1, j\neq k}^K\bar{\pmb\phi}_j^{\dag}\bar{\mathbf h}_{j,k}\bar{\mathbf h}_{j,k}^{\dag}\bar{\pmb\phi}_j}\label{eq:5}\\
\text{s.t.~}&~ \sum_{m=1}^M\alpha||\bar{\pmb\Phi}^m||_2\leq \delta,\label{eq:6}\\
&~\bar{\pmb\phi}_k^{\dag}{\mathbf e}_n{\mathbf e}_n^{\dag}\bar{\pmb\phi}_k\leq p_k^{\max}, \forall k\in{\cal K}; n=1, 2, \ldots, N, \label{eq:7}
\end{align}
where ${\mathbf e}_n\in{\mathbb R}^{N\times 1}$ is an elementary vector with a one at the $n\text{-th}$ position,
The weight $\alpha$ can be regarded as a free parameter in the convex relaxation, whose value can be chosen to avoid trivial solutions (i.e, $\bar{\pmb\Phi}^m={\mathbf 0}$ or $\bar{\pmb\Phi}^m\neq {\mathbf 0}, \forall m\in{\cal M}$).
Notably, in (\ref{eq:6}),  both $\alpha$ and $\delta$ affect the cardinal number of the triggered module subsets together.
%Obviously, in (\ref{add:3}), both $\alpha$ and $\delta$ together affect the sparsity of IRS.
The sparsity is negatively correlated with $\alpha$, while is positively correlated with  $\delta.$
Inspired by the iterative algorithm of redefining the weights \cite{Candes2008Enhancing}, intuitively, parameter $\alpha$ should relate inversely to $\delta,$ consequently,  $\alpha$ can be  set to $1/(\delta+0.01)$.
Our motivation for introducing $0.01$ in the $\alpha$ setting is to provide stability and ensure feasibility.
In this setting, the number of activated modules increases as the parameter $\delta$ increases until approaching the upper bound of the module quantity.
In addition, for a given system setting (i.e., $M, K, N,$ and $\{p_k^{\max}\}$), there is a rang of favorable parameter $\delta$, i.e.,  $(0, -0.005+0.5\sqrt{(0.01)^2+\sqrt{16MKN\max_{k}\{p_k^{\max}\}}}).$

By introducing an auxiliary variable $\gamma,$ the joint activated modules identification, transmit power allocation, and the corresponding passive beamformer design problem (P1) can thus be equivalent to
\begin{align}
\text{(P2)}~&\max_{\bar{\pmb\Phi}, \gamma}~\gamma  \label{eq:8}\\
\text{s.t.~}&~\frac{\bar{\pmb\phi}_k^{\dag}\bar{\mathbf h}_{k,k}\bar{\mathbf h}_{k,k}^{\dag}\bar{\pmb\phi}_k}
{\sigma^2+\sum_{j=1, j\neq k}^K\bar{\pmb\phi}_j^{\dag}\bar{\mathbf h}_{j,k}\bar{\mathbf h}_{j,k}^{\dag}\bar{\pmb\phi}_j}\geq \gamma, \label{eq:9}\\
&~~ (\ref{eq:6})~\text{and}~ (\ref{eq:7}).\label{eq:10}
\end{align}
Clearly, for large $\gamma,$ problem (P2) may be infeasible due to  the resulting stringent SINR constraints, strong interference, and insufficient number of activated  modules.
Thus, in the following, problem (P2) can be solved efficiently via bisection method for feasibility checking.

\section{Activated Modules  Identification and the Minimum SINR  Maximization }

\subsection{Activated Modules Identification}
As mentioned early,  for a given $\gamma>0,$ the design problem (P2) becomes the feasibility test one.
In this context, the challenge in solving problem (P2) lies in the fact that its objective is non-differentiable and that the feasible set is nonconvex.
To proceed further, for the fixed $\gamma$, we observe that (P2) is feasible if and only if the solution of the following optimization problem (P3) is lower than $\delta,$ where (P3) is given by
\begin{align}
\text{(P3)}~ &\min_{\bar{\pmb\Phi}} ~ \sum_{m=1}^M\alpha||\bar{\pmb\Phi}^m||_2\label{eq:12}\\
\text{s.t.~}&\sqrt{(1+\gamma^{-1})}\bar{\mathbf h}_{k,k}^{\dag}\bar{\pmb\phi}_k\geq ||[\bar{\mathbf H}^{k\dag}\widetilde{\pmb\Phi}, \sigma] ||_2,\label{eq:13}\\
&\bar{\pmb\phi}_k^{\dag}{\mathbf e}_n{\mathbf e}_n^{\dag}\bar{\pmb\phi}_k\leq p_k^{\max},\label{eq:14}\\
&\text{Im}(\bar{\mathbf h}_{k,k}^{\dag}\bar{\pmb\phi}_k)=0, \forall k\in{\cal K}. \label{eq:15}
\end{align}
The constraint (\ref{eq:13}) of (P3) is the reformulation of SINR constraint (\ref{eq:9}) relies on the second-order cone program \cite{Boyd2009Convex}, where the $(NK)\times K$ matrix $\tilde{\pmb\Phi}$ is defined as $\tilde{\pmb\Phi}=\text{diag}\{\bar{\pmb\phi}_1, \bar{\pmb\phi}_2,\ldots, \bar{\pmb\phi}_K\}.$
As problem (P3) is a convex  problem, it can be optimally solved by existing
convex optimization solvers such as CVX.
To this end,  for a special $\gamma,$ the feasibility of problem (P2) can be checked by solving the convex problem (P3).
In each step, we check the feasibility of (P2) for a specific $\gamma.$
As the bisection procedure converges, the activated modules can be determined directly by
exploring the row block sparse pattern of  $\tilde{\pmb\Phi}.$
Once the activated modules are identified, we drop the module size constraint and solve the conventional max-min problems to get the transmit power and the passive beamforming of the activated modules.
Alternating optimization technique as an established tool for max-min SINR problems of this kind, which will be used in the following section.
%Alternating optimization technique as the established theoretical tool for max-min problems in this kind of has been used in many articles \cite{ Wu2018Intelligent,Guo2019Weighted,  Bjornson2019Intelligent, Huang2019Reconfigurable, Zenzo2019Reconfigurable} will be used in the following.
%Thus, the detailed process of obtaining the optimal transmit power and passive beamforming will not be given.
\subsection{Minimum SINR Maximization}
In this section, we solve joint transmit power allocation and passive beamformer design when the activated modules are identified.
To be specific,  for the original max-min SINR problem (P0),  the module size constraint is dropped and the diagonal blocks of $\pmb\Phi$ corresponding to the non-activated modules are forced to be zero.
For convenience to illustrate, the phase-shift matrix with identified triggered modules denoted by ${\cal F}_{\pmb\Phi}.$
Particularly, we focus on solving:
\begin{equation}\label{eq:20}
\begin{aligned}
&\max_{\{p_k\}_{k\in{\cal K}}, {\cal F}_{\pmb\Phi}}\min_{k\in{\cal K}} \frac{p_k|{\mathbf g}_k^{\dag}{\cal F}_{\pmb\Phi}{\mathbf h}_k|^2}
{\sum_{j=1,j\neq k}^Kp_j|{\mathbf g}_k^{\dag}{\cal F}_{\pmb\Phi}{\mathbf h}_j|^2+\sigma^2}\\
&\text{s.t.~}(\ref{eq:3-2})  \text{~and~} (\ref{eq:3-3}).
\end{aligned}
\end{equation}
Note that (P0) can be transformed into the conventional max-min SINR problem, which can be efficiently and optimally solved by employing the alternating optimization technique \cite{Hestenes1969Multiplier} to separately and iteratively solve for $\{p_k\}_{k\in{\cal K}}$ and ${\cal F}_{\pmb\Phi}.$
In the rest of this section, the optimization with respect to ${\cal F}_{\pmb\Phi}$ for fixed $\{p_k\}_{k\in{\cal K}}$, and with respect to $\{p_k\}_{k\in{\cal K}}$ for fixed ${\cal F}_{\pmb\Phi}$ will be treated separately.

\subsubsection{Optimizing Phase-Shift Matrix ${\cal F}_{\pmb\Phi}$}
Let ${\cal F}_{\pmb\phi}\in{\mathbb C}^{N\times 1}$ denote the vectorization of diagonal matrix ${\cal F}_{\pmb\Phi}.$ Substituting $\bar{\mathbf h}_{j,k}$ into the objective function of (\ref{eq:20}), then, ${p_k|{\mathbf g}_k^{\dag}{\cal F}_{\pmb\Phi}{\mathbf h}_k|^2}=p_k{{\cal F}_{\pmb\phi}}^{\dag}\bar{\mathbf h}_{k,k}\bar{\mathbf h}_{k,k}^{\dag}{\cal F}_{\pmb\phi},$
${\sum_{j=1,j\neq k}^Kp_j|{\mathbf g}_k^{\dag}{\cal F}_{\pmb\Phi}{\mathbf h}_j|^2+\sigma^2}=\sum_{j=1,j\neq k}^{K}p_j{\cal F}_{\pmb\phi}^{\dag}\bar{\mathbf h}_{j,k}\bar{\mathbf h}_{j,k}^{\dag}{\cal F}_{\pmb\phi}+\sigma^2$ for all $k$ and $j.$
Utilizing the method of \textit{partial linearization for generalized fractional programs} \cite{Benadada1988Partial}, introducing the continuous functions w.r.t. ${\cal F}_{\pmb\phi}$, are defined as
\begin{align}
u_k({\cal F}_{\pmb\phi})&=p_k{\cal F}_{\pmb\phi}^{\dag}\bar{\mathbf h}_{k,k}\bar{\mathbf h}_{k,k}^{\dag}{\cal F}_{\pmb\phi}\\
v_k({\cal F}_{\pmb\phi})&=\sum_{j=1, j\neq k}^K p_j{\cal F}_{\pmb\phi}^{\dag}\bar{\mathbf h}_{j,k}\bar{\mathbf h}_{j,k}^{\dag}{\cal F}_{\pmb\phi}+\sigma^2.
\end{align}
By introducing parameter $\gamma_{\text{out}}$, then the optimization with respect to phase-shift matrix is equivalent to
\begin{align}
\max_{{\cal F}_{\pmb\phi}, \gamma_{\text{out}}} &~ \gamma_{\text{out}}\\
\text{s.t.~}&~ u_k({\cal F}_{\pmb\phi})-\gamma_{\text{out}}v_k({\cal F}_{\pmb\phi})\geq 0, \text{~and~ } (\ref{eq:3-4}).
\end{align}
For $1\leq k\leq K,$ denote ${\cal G}_k({\cal F}_{\pmb\phi}, \gamma_{\text{out}})=u_k({\cal F}_{\pmb\phi})-\gamma_{\text{out}}v_k({\cal F}_{\pmb\phi})$, and consider the following partial linearization of ${\cal G}_k({\cal F}_{\pmb\phi}, \gamma_{\text{out}})$ at a point $({\cal F}_{\pmb\phi}^{(\tau)}, \gamma_{\text{out}}^{(\tau)})$
\begin{equation}
\begin{aligned}
{\cal G}_k^{(\tau)}&({\cal F}_{\pmb\phi},\gamma_{\text{out}})\\
&={\cal G}_{k}({\cal F}_{\pmb\phi}, \gamma_{\text{out}}^{(\tau)})+(\gamma_{\text{out}}-\gamma_{\text{out}}^{(\tau)})\nabla_{\gamma_{\text{out}}}{\cal G}({\cal F}_{\pmb\phi}^{(\tau)}, \gamma_{\text{out}}^{(\tau)})\\
&=u_k({\cal F}_{\pmb\phi})-\gamma_{\text{out}}^{(\tau)}v_k({\cal F}_{\pmb\phi})-(\gamma_{\text{out}}-\gamma_{\text{out}}^{(\tau)})v_k({\cal F}_{\pmb\phi}^{(\tau)}).
\end{aligned}
\end{equation}
%which can be efficiently and optimally solved by generalized fractional programs \cite{Crouzeix1985An, Borde1987Convergence, Benadada1988Partial}.
The following sub-problem specified with this partial linearization of the ${\cal G}_k$ is to be solved by CVX at each iteration $\tau$ of the algorithm.
\begin{equation}\label{eq:21}
\begin{aligned}
\max_{\gamma_{\text{out}}}~ &\gamma_{\text{out}}\\
\text{s. t. ~}&~{\cal G}_k^{(l)}({\cal F}_{\pmb\phi}, \gamma_{\text{out}})\geq 0, \forall k=1, 2, \ldots, K; \text{~and~} (\ref{eq:3-4}).
\end{aligned}
\end{equation}
\subsubsection{Optimization with Respect to the Power Allocation $\{p_k\}_{k\in{\cal K}}$}
Likewise, for the case where ${\cal F}_{\pmb\Phi}$ is fixed and the objective is the optimization over ${\mathbf p}=[p_1, p_2, \ldots, p_K]^T,$ we introduce continuous functions of ${\mathbf p}$, denoted by $\xi_k({\mathbf p})$ and $\eta_k({\mathbf p}),$ respectively, and are defined as
\begin{align}
\xi_k({\mathbf p})&=p_k|{\mathbf g}_k^{\dag}{\cal F}_{\pmb\Phi}{\mathbf h}_k|^2, \forall k=1,2, \ldots, K\\
\eta_k(\mathbf p)&=\sum_{j=1,j\neq k}^K p_j|{\mathbf g}_k^{\dag}{\cal F}_{\pmb\Phi}{\mathbf h}_j|^2+\sigma^2,\forall k=1, 2, \ldots, K.
\end{align}
Denote $\Omega_k({\mathbf p}, \gamma_{\text{out}})=\xi_k({\mathbf p})-\gamma_{\text{out}}\eta_k({\mathbf p}).$
The partial linearalization of $\Omega_k$ at a point $({\mathbf p}^{(\tau)}, \gamma_{\text{out}}^{(\tau)})$ is
\begin{align}
\Omega_k^{(\tau)}({\mathbf p}, \gamma_{\text{out}})&=\Omega_k({\mathbf p}, \gamma_{\text{out}}^{(\tau)})+(\gamma_{\text{out}}
-\gamma_{\text{out}}^{(\tau)})\nabla_{\gamma_{\text{out}}}\Omega_k({\mathbf p}^{(l)}, \gamma_{\text{out}}^{(\tau)})\\
&=\xi_k({\mathbf p})-\gamma_{\text{out}}^{(\tau)}\eta_k({\mathbf p})-(\gamma_{\text{out}}-\gamma_{\text{out}}^{(\tau)})\eta_k({\mathbf p}^{(\tau)}).
\end{align}
Consequently, we focus on solving:
\begin{equation}\label{eq:22}
\begin{aligned}
\max_{{\mathbf p}, \gamma_{\text{out}}}~ &~\gamma_{\text{out}}\\
\text{s.t.~}&~\Omega_k^{(\tau)}({\mathbf p}, \gamma_{\text{out}})\geq 0, \text{~and ~} (\ref{eq:3-2}).
\end{aligned}
\end{equation}

In the proposed alternating optimization algorithm, we solve ${\mathbf p}$ and ${\cal F}_{\pmb\phi}$ by addressing problems (\ref{eq:21}) and (\ref{eq:22}) alternately in an iterative manner, where the solution obtained in each iteration
is used as the initial point of the next iteration. The details of the proposed algorithm are
summarized in Algorithm 1.
\begin{algorithm}[!htp]%\label{algorithm:1}         %Ëã·¨µÄ¿ªÊ¼
\caption{ Alternating optimization algorithm for ${\mathbf p}$ and ${\cal F}_{\pmb\phi}$ when the trigger modules are identified}             % Ëã·¨µÄ±êÌâ
\label{alg:Framwork}
\textit{ Step 0:} Initialize  ${\cal F}_{\pmb\phi}^{(0)} $ and ${\mathbf p}^{(0)}$ to feasible values, and let $\gamma_{\text{out}}^{(0)}=\min_{1\leq k\leq K}\left\{\frac{u_k({\cal F}_{\pmb\phi}^{(0)})}{v_k({\cal F}_{\pmb\phi}^{(0)})}  \right\}, $
and set the iteration number $\tau=0.$

{\bf repeat}

{\textit{Step 1:}  Solve (\ref{eq:21}) by CVX for given ${\mathbf p}^{(\tau)},$
and denote ${\cal F}_{\pmb\phi}^{(\tau)}$ be an optimal solution. }

{\textit{Step 2:} Solve problem (\ref{eq:22}) for given ${\cal F}_{\pmb\phi}^{(\tau)},$ and denote the optimal solution as ${\mathbf p}^{(\tau+1)}$.

}

{\textit{Step 3: } Update $\tau=\tau+1$.}

{\textit{Step 4.}  {\bf until}  $\gamma_{\text{out}}$ converges or problem becomes infeasible. }

\end{algorithm}
\begin{figure*}[!t]
	\centering
	\subfigure{
		\begin{minipage}[t]{0.23\linewidth}
			\includegraphics[width=1.8in]{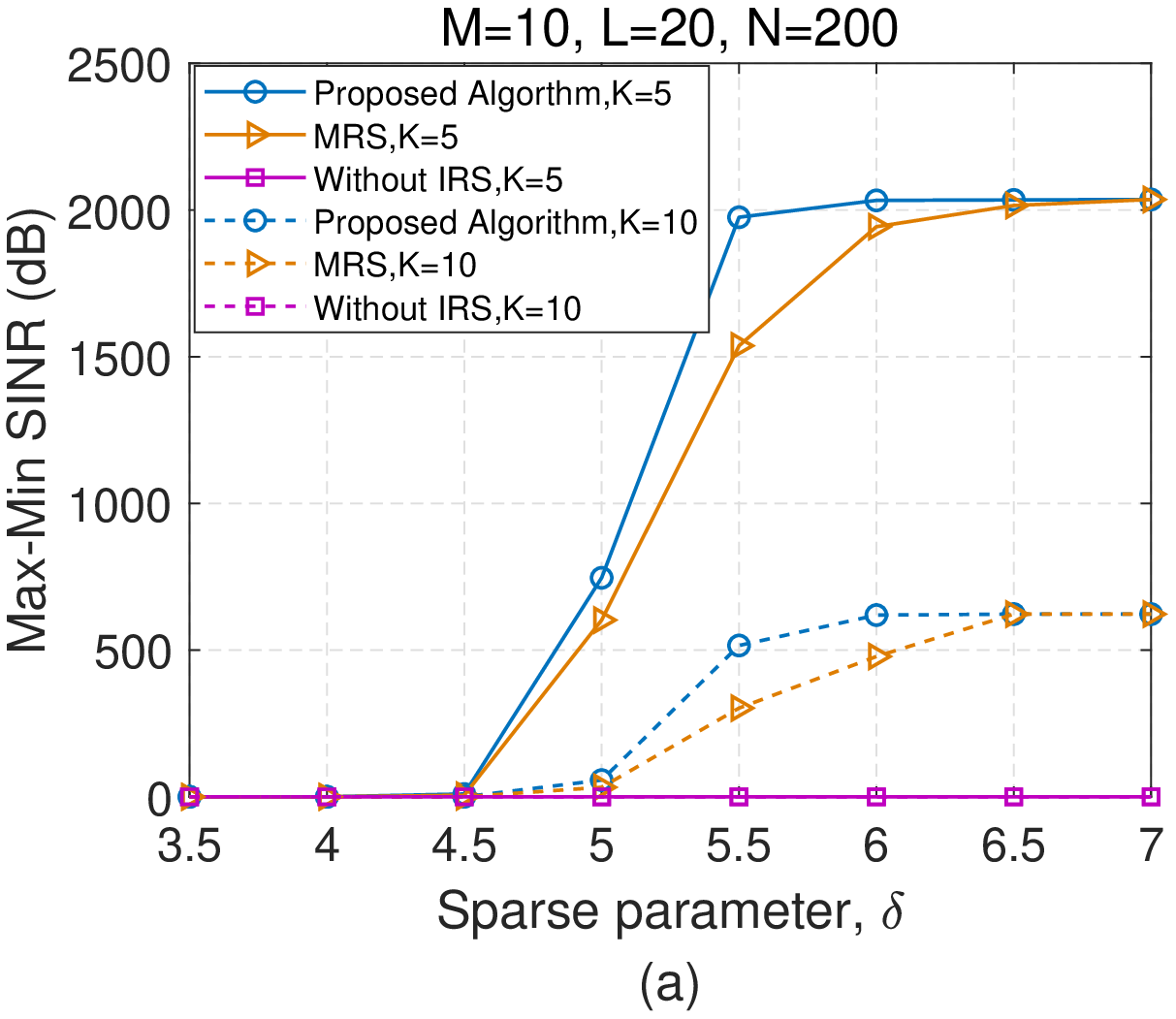}
		\end{minipage}
		\label{fig:1-1}
	}
	\centering
	\subfigure{
		\begin{minipage}[t]{0.23\linewidth}
			\includegraphics[width=1.8in]{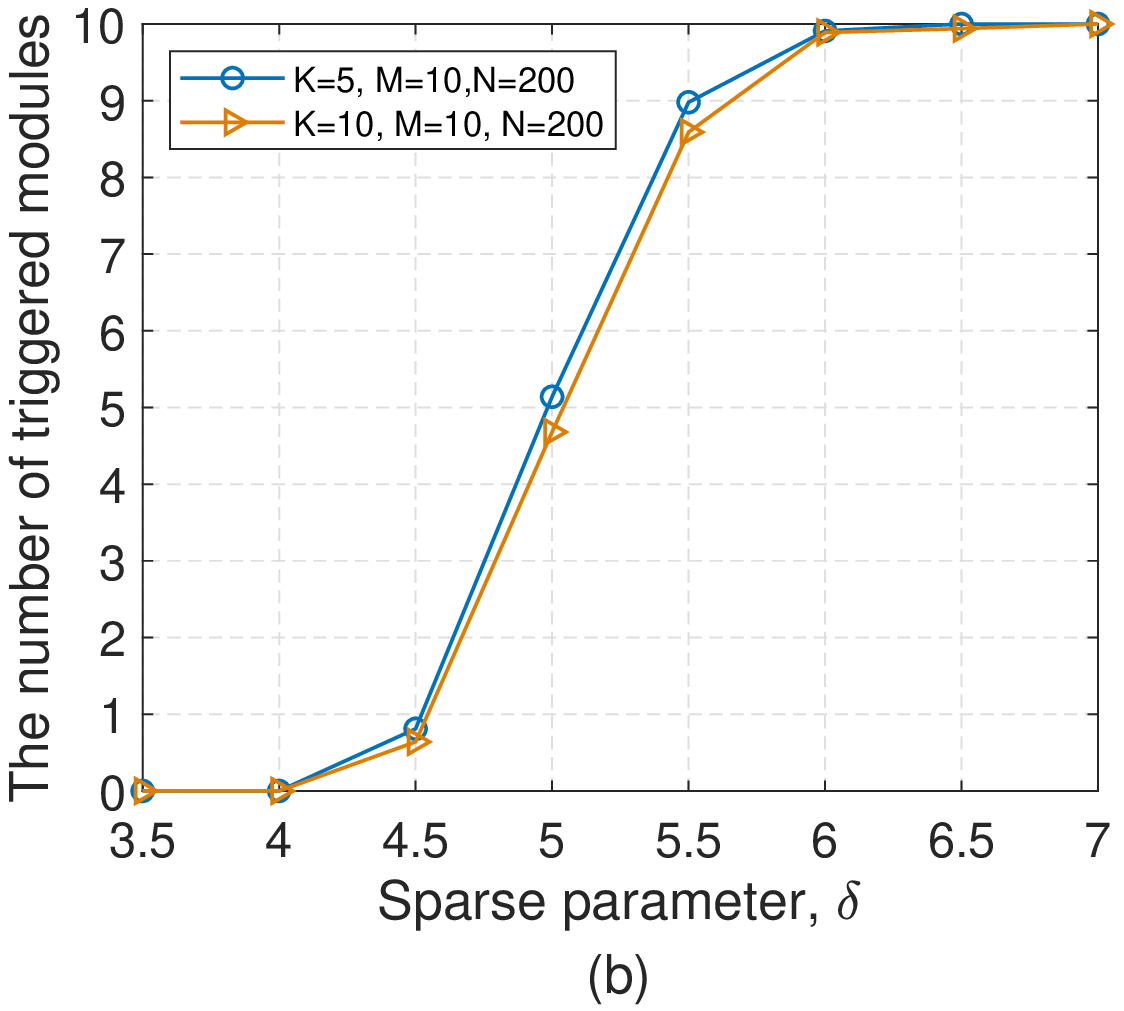}
		\end{minipage}
		\label{fig:1-2}
	}
	\centering
	\subfigure{
		\begin{minipage}[t]{0.23\linewidth}
			\includegraphics[width=1.8in]{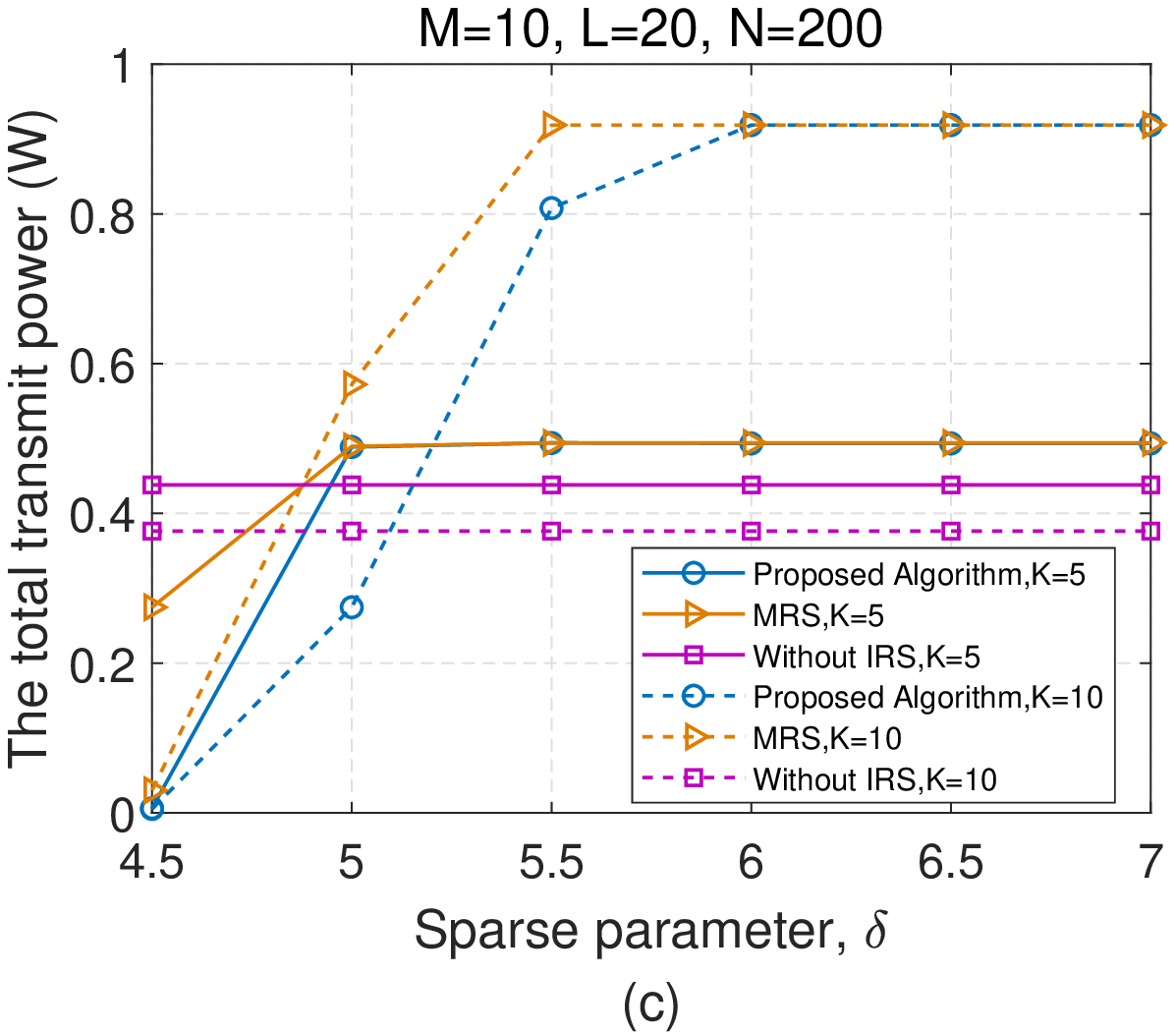}
		\end{minipage}
		\label{fig:1-3}
	}
	\centering
	\subfigure{
		\begin{minipage}[t]{0.23\linewidth}
			\includegraphics[width=1.8in]{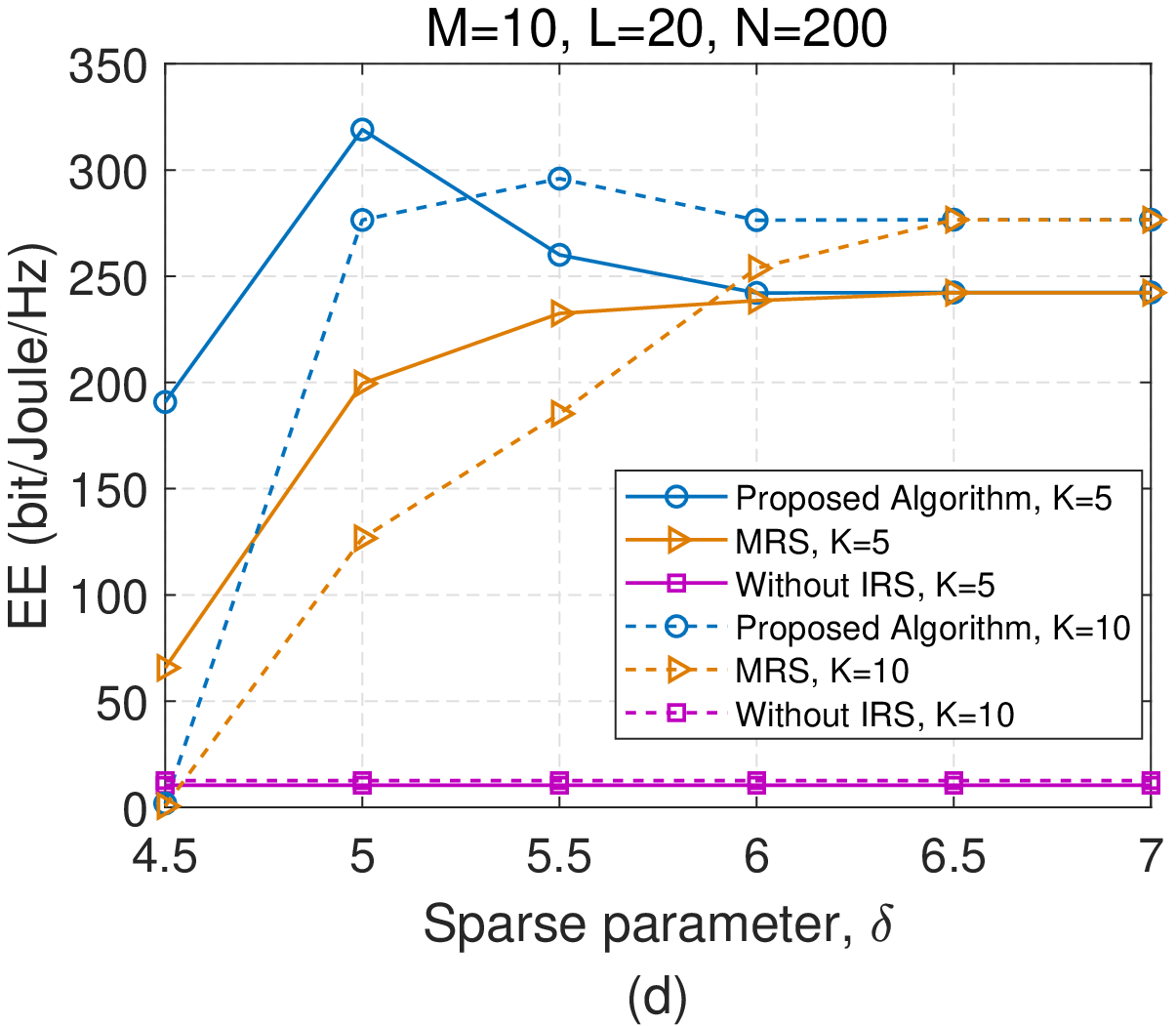}
		\end{minipage}
		\label{fig:1-4}
	}
	\caption{ (a) Max-min SINR, (b) the number of triggered modules, (c) the total transmit power, and (d) EE versus the sparse parameter $\delta$ using $K=5, 10$, for $M=10$ and $p^{\max}=20\text{~dBm}.$}
	\label{fig:1}
\end{figure*}
\section{Simulation Results}
We evaluate the performance of the proposed joint design of activated module subset identification, transmit power allocation, and the corresponding passive beamformer in the IRS-aided cooperative communication networks.
The convergence property and effectiveness of the two-block ADMM algorithm are verified.
We consider the IRS-aided cooperative communication system consisting with $K$ ST-DT pairs and an IRS with $N=ML$ reflecting elements, where $M$ is the number of IRS modules and $L$ is the number of reflecting elements of each module.
Suppose that the $K$ STs are randomly and uniformly deployed within a circle cell centered at $(0,0)~\text{m}$ with the cell radius $2~\text{m}$,
	and the corresponding $K$ DTs are located within a circle cell with radius $2~\text{m}$ centered at $(200, 0)~\text{m}.$
	Besides, the IRS is assumed to be fixed at the location $(120, 50)\text{~m}.$
	Unless specified otherwise, the simulation setting  is given as follows.
	The number of ST-DT pairs is less than or equal to the number of modules at the IRS, i.e, $K\leq M.$
	We consider an IRS-aided communication system with carrier frequency  $2.3\text{~GHz}$ and a system bandwidth ${\cal W}=10\text{~MHz}.$
	From \cite{Zheng2020Intelligent}, we set the path loss exponent of the ST-DT pair direct link as $3.5,$   and the path loss at the reference distance  $1\text{~m}$ is set as $30\text{~dB}$ for each individual link \cite{Wu2019Intelligentjournal}.
	For the IRS-aided link, $2$ and $2.1$ are the values of the path loss exponents from STs to the IRS and that from the IRS to DTs, respectively.
	Moreover, the path loss model for the NLOS paths is characterized by Rayleigh fading.
	Channel vectors $\{{\mathbf h}_{k}\}$ and $\{{\mathbf g}_{k}\}$ are  generated as i.i.d. zero-mean complex Gaussian random vectors, where the variance of each channel is determined using pathloss model $\sigma_{{\mathbf h}_{k}}^2=(200/d_{{\mathbf h}_{k}})^2$ with $d_{{\mathbf h}_{k}}$ as the distance between ST $k$ and IRS \cite{Lin2016Joint}.
	Likewise, $\{{\mathbf g}_{k}\}$ can be generated according to the distribution ${\cal C}{\cal N}(0,\sigma_{{\mathbf g}_{k}}^2)$, where the variance is given by $\sigma_{{\mathbf g}_{k}}^2=(200/d_{{\mathbf g}_{k}})^{2.1}$ with $d_{{\mathbf g}_{k}}$ being the distance between IRS and DT $k.$
	%Channel vectors $\{{\mathbf h}_k\}$ and $\{{\mathbf g}_k\}$ are generated as i.i.d. zero-mean complex Gaussian random vectors, where the variance of each channel is
	We assume quasi-static block fading channels in this paper, i.e., the channels from the STs to the IRS and that from the IRS to the DTs remain constant during each time block, but may vary from one to another \cite{Yang2019Low}.
%	For the proposed algorithm, the convergence tolerance is $\epsilon=10^{-4}.$
	For simplicity, all the STs are assumed to have the same maximum transmit power, i.e., $p_k^{\max}=p^{\max}=20\text{~dBm}$ and the noise power at all the destination terminals is assumed to be identical with $\sigma^2=-90\text{~dBm}.$
	The number of reflecting elements of each module  is $L=20$  and each user terminal is equipped with a single antenna.
	All the simulation results are obtained by averaging over $10^4$ channel realizations.
	%Throughout the simulations, unless otherwise specified, we adopt the parameters reported in Table \ref{tab:1} (see \cite{Huang2019Reconfigurable,Zheng2020Intelligent} and references therein).

For IRS-aided communication systems, we evaluate the performance of the proposed algorithm with two baseline schemes in our simulations. For baseline 1 (i.e., the case without IRS), only the S-D direct link is considered where the number of reflecting elements at IRS is set as $N=0.$ Baseline 2 denoted as the method of randomly selecting the activated modules (MRS).Besides, to draw more insight for the superiority of the introduced modular IRS structure, we further compare the energy efficiency (EE) performance, where $\texttt{EE}$ (bit/Joule/Hz) is defined as the ratio of the network achievable sum rate and the overall power consumption, i.e., $\texttt{EE}=\frac{\sum_{k=1}^K R_k}{P_{\text{total}}}.$
Inspired by \cite{Huang2019Reconfigurable}, the overall power consumption of the IRS-aided system can be expressed as \begin{equation}
\begin{aligned}
P_{\text{total}}^{\text{IRS}}=&\xi_{\text{ST}}\sum_{k=1}^K p_k+KP_{\text{ST}}+KP_{\text{DT}}\\
&+card(\text{subset of triggered modules})\cdot P(L),
\end{aligned}
\end{equation}
where $P_{\text{ST}}$ and $P_{\text{DT}}$ denote the hardware static power dissipated by each ST and DT, respectively, $\xi_{\text{ST}}$ the circuit dissipated power coefficient at each ST, and $P(L)$ is the power consumption of each module having $L$ reflecting elements. Moreover, let $P_{\text{ST}}=P_{\text{DT}}=10\text{~dBm}$, $P(L)=(L\cdot 0.01)\text{W}$, and $\xi_{\text{ST}}=1.2.$

Figures \ref{fig:1-1} and \ref{fig:1-2} illustrate the effects of the number of S-D pairs $K$ on the SINR performance and the number of triggered modules, respectively, based on all above schemes.
Two simulation cases with $K=5$ and $K=10$ are shown with the same number of modules $M=10$ at the IRS and the maximum transmit power of each ST is $p^{\max}=20\text{~dBm}.$
From the results, we observe that the SINR achieved by all above schemes besides the baseline 1 first increases and then remain constant, when $\delta$ increases.
Apparently, the reduction of SINR for all above schemes besides the baseline 1 due to increasing $K$ is significant.
This in essence attributes to that more interference will be induced from concurrent transmissions if the IRS serves for more ST-DT pairs.
As seen in Fig.\ref{fig:1-2}, for given values of $M$ and $p^{\max}$, a sparser solution will be achieved for $K=10$ than that of $K=5.$
This is due to the fact that the module size constraint is more stringent with $K=5$ than that of $K=10$ for the same value of $\delta,$ based on the valid range of parameter $\delta$.
In addition to our observations in Fig. \ref{fig:1-1} and \ref{fig:1-2} with respect to SINR and the number of activated modules, in Fig. \ref{fig:1-3}, we evaluate and compare the total transmit power versus the sparse parameter using different number of user pairs for a given value of $M.$
As expected,  for $K\in\{5, 10\},$  the total transmit power by the proposed algorithm is lower than that of MRS for a given value of $\delta.$

To measure the benefits of the proposed modular activation mechanism with respect to the existing full activation setting, Fig. \ref{fig:1-4} depicts the EE achieved by all mentioned schemes versus sparse parameter for $M=10.$
As seen in Fig. \ref{fig:1-4}, for $\delta<6$ with $K=5$ ($\delta<6.5$ with $K=10$), the proposed  algorithm in IRS-aided communication significantly outperforms MRS.
It is interesting to notice that, for simulation settings $K=5$ and $K=10$, the EE achieved by the proposed algorithm first increases and then decreases until to a saturation value, when the value of $\delta$ increases.
The reason is that when $\delta$ relatively small, e.g., $\delta\in[4.5, 5]$ for $K=5$ ($\delta\in[4.5, 5.5]$ for $K=10$), the increase of SINR dominates the maximizing the EE of system in this regime.
By contrast, as the value of sparse parameter becomes is larger than the optimal $\delta,$ e.g., $\delta>5$ for $K=5$ ($\delta>5.5$ for $K=10$), more and more modules are activated for cooperative communication, consequently, the circuit power consumption dominates the total power consumption rather than the transmit power consumption.
Therefore, for any given network setting, there is an optimal choice of $\delta,$ which leads to the cost-effective reflecting element schedule.

\section{Conclusions}
In this paper, we studied the joint problem of RRA and both transmit power allocation and passive beamformer design to IRS-aided cooperative communication networks.
Specifically, the RRA can be realized via  module activation, which is based on the introduced modular IRS structure, which contains multiple modules, each module is attached with a smart controller.
Our goal was to maximize the minimum SINR at DTs via joint modules activation and active-passive beamformer design while satisfying the transmit power at each ST.
In order to obtain the low-complexity solution, we developed the approximate convex problem of the max-min problem via convex relaxation to the module size constraint.
Simulation results showed the meaningfulness of the introduced modular IRS structure.

%\bibliographystyle{ieeetr}
%\bibliography{newbibglobecom}

\section*{Acknowledgement}
{This work was supported in part by the National Science Foundation of China under Grant number 61671131, also supported by the National Research Foundation (NRF), Singapore, under Singapore Energy Market Authority (EMA), Energy Resilience, NRF2017EWT-EP003-041, Singapore NRF2015-NRF-ISF001-2277, Singapore NRF National Satellite of Excellence, Design Science and Technology for Secure Critical Infrastructure NSoE DeST-SCI2019-0007, A*STAR-NTU-SUTD Joint Research Grant on Artificial Intelligence for the Future of Manufacturing RGANS1906, Wallenberg AI, Autonomous Systems and Software Program and Nanyang Technological University (WASP/NTU) under grant M4082187 (4080), Singapore Ministry of Education (MOE) Tier 1 (RG16/20), and NTU-WeBank JRI (NWJ-2020-004), Alibaba Group through Alibaba Innovative Research (AIR) Program, Alibaba-NTU Singapore Joint Research Institute (JRI),
	Nanyang Technological University (NTU) Startup Grant, Alibaba-NTU Singapore Joint Research Institute (JRI),
	Singapore Ministry of Education Academic Research Fund Tier 1 RG128/18, Tier 1 RG115/19, Tier 1 RT07/19, Tier 1 RT01/19, and Tier 2 MOE2019-T2-1-176,
	NTU-WASP Joint Project,
	Singapore National Research Foundation (NRF) under its Strategic Capability Research Centres Funding Initiative: Strategic Centre for Research in Privacy-Preserving Technologies \& Systems (SCRIPTS),  
	Energy Research Institute @NTU (ERIAN),
	Singapore NRF National Satellite of Excellence,
	Design Science and Technology for Secure Critical Infrastructure NSoE DeST-SCI2019-0012,
	AI Singapore (AISG) 100 Experiments (100E) programme,
	NTU Project for Large Vertical Take-Off \& Landing (VTOL) Research Platform.}

\end{document}